\documentstyle[12pt]{article}
\expandafter\chardef\csname pre amssym.def at\endcsname=\the\catcode`\@
\catcode`\@=11
\def\undefine#1{\let#1\undefined}
\def\newsymbol#1#2#3#4#5{\let\next@\relax
 \ifnum#2=\@ne\let\next@\msafam@\else
 \ifnum#2=\tw@\let\next@\msbfam@\fi\fi
 \mathchardef#1="#3\next@#4#5}
\def\mathhexbox@#1#2#3{\relax
 \ifmmode\mathpalette{}{\m@th\mathchar"#1#2#3}%
 \else\leavevmode\hbox{$\m@th\mathchar"#1#2#3$}\fi}
\def\hexnumber@#1{\ifcase#1 0\or 1\or 2\or 3\or 4\or 5\or 6\or 7\or 8\or
 9\or A\or B\or C\or D\or E\or F\fi}

\font\tenmsa=msam10
\font\sevenmsa=msam7
\font\fivemsa=msam5
\newfam\msafam
\textfont\msafam=\tenmsa
\scriptfont\msafam=\sevenmsa
\scriptscriptfont\msafam=\fivemsa
\edef\msafam@{\hexnumber@\msafam}
\mathchardef\dabar@"0\msafam@39
\def\dashrightarrow{\mathrel{\dabar@\dabar@\mathchar"0\msafam@4B}}
\def\dashleftarrow{\mathrel{\mathchar"0\msafam@4C\dabar@\dabar@}}

\def\ulcorner{\delimiter"4\msafam@70\msafam@70 }
\def\urcorner{\delimiter"5\msafam@71\msafam@71 }
\def\llcorner{\delimiter"4\msafam@78\msafam@78 }
\def\lrcorner{\delimiter"5\msafam@79\msafam@79 }
\def\yen{{\mathhexbox@\msafam@55 }}
\def\checkmark{{\mathhexbox@\msafam@58 }}
\def\circledR{{\mathhexbox@\msafam@72 }}
\def\maltese{{\mathhexbox@\msafam@7A }}

\font\tenmsb=msbm10
\font\sevenmsb=msbm7
\font\fivemsb=msbm5
\newfam\msbfam
\textfont\msbfam=\tenmsb
\scriptfont\msbfam=\sevenmsb
\scriptscriptfont\msbfam=\fivemsb
\edef\msbfam@{\hexnumber@\msbfam}
\def\Bbb#1{\fam\msbfam\relax#1}
\def\widehat#1{\setboxz@h{$\m@th#1$}%
 \ifdim\wdz@>\tw@ em\mathaccent"0\msbfam@5B{#1}%
 \else\mathaccent"0362{#1}\fi}
\font\teneufm=eufm10
\font\seveneufm=eufm7
\font\fiveeufm=eufm5
\newfam\eufmfam
\textfont\eufmfam=\teneufm
\scriptfont\eufmfam=\seveneufm
\scriptscriptfont\eufmfam=\fiveeufm
\def\frak#1{{\fam\eufmfam\relax#1}}

\catcode`\@=\csname pre amssym.def at\endcsname

\undefine\hbar
\newsymbol\ltimes 226E
\newsymbol\upharpoonright 1316
\newsymbol\hbar 207E
\newsymbol\hbar 207E
\def\Box{\hbox{\vrule height1ex\kern-0.4pt
\vbox to 1ex{\hrule width1ex\vfil\hrule width1ex}\kern-0.4pt\vrule height1ex}}
\newcommand{\sqr}[2]{{{\vcenter{\vbox{\hrule height.#2pt
\hbox{\vrule width.#2pt height#1pt \kern#1pt
\vrule width.#2pt}
\hrule height.#2pt}}}}}

\topmargin = - 0.5 cm
\newcommand{\til}{\tilde}
\newtheorem{theorem}{Theorem}

\newtheorem{defn}{Definition}

\newtheorem{con}{Conjecture}

\newcommand{\Oh}{\Omega_{\hbar}}
\newtheorem{prop}{Proposition}
\newcommand{\wi}{W^{\hbar}}

\newcommand{\cci}{C_c^{\infty}}
\newcommand{\cin}{C^{\infty}}

\newcommand{\eo}{\setcounter{equation}{0}}

\newcommand{\be}{\begin{equation}}

\newcommand{\ee}{\end{equation}}
\newcommand{\ca}{$C^*$-algebra\mbox{ }}

\newcommand{\spe}{{\rm spec}}
\newcommand{\na}{\partial}
\newcommand{\al}{\alpha}

\newcommand{\gm}{\gamma}
\newcommand{\Gm}{\Gamma}
\newcommand{\Rn}{{\Bbb R}^n}
\newcommand{\dl}{\delta}
\newcommand{\Dl}{\Delta}

\newcommand{\th}{\theta}

\newcommand{\lm}{\lambda}

\newcommand{\sg}{\sigma}
\newcommand{\ta}{\tau}
\newcommand{\ph}{\phi}
\newcommand{\phv}{\varphi}
\newcommand{\ch}{\chi}
\newcommand{\ovl}{\overline}
\newcommand{\ah}{\alpha_{\hbar}}
\newcommand{\sh}{\sigma_{\hbar}}
\newcommand{\ps}{\psi}
\newcommand{\om}{\omega}
\newcommand{\Om}{\Omega}
\newcommand{\nn}{\nonumber}

\newcommand{\raw}{\rightarrow}

\newcommand{\half}{\mbox{\footnotesize $\frac{1}{2}$}}
\newcommand{\A}{{\frak A}}

\newcommand{\bib}{\bibitem}

\renewcommand{\H}{{\cal H}}
\newcommand{\KH}{{\cal K(H)}}

\newcommand{\gr}{groupoid\mbox{ }}

\newcommand{\Qh}{Q_{\hbar}}
\newcommand{\g}{\mbox{\bf g}}

\newcommand{\ff}{\check{f}}
\newcommand{\xd}{\dot{x}}

\newcommand{\xm}{x^{\mu}}

\newcommand{\pmu}{p_{\mu}}

\newcommand{\lpl}{\lim_{\hbar\rightarrow 0}}

\newcommand{\h}{\mbox{\bf h}}

\newcommand{\n}{\parallel}

 \renewcommand{\ll}{\label}

\renewcommand{\O}{{\cal O}}
\newcommand{\K}{{\cal K}}

\newcommand{\pcs}{\pi_c^S}
\newcommand{\pqs}{\pi_q^{\chi}}
\newcommand{\Hs}{{\cal H}_{\chi}}

\newcommand{\Ah}{{\cal A}_{\hbar}}

\newcommand{\pl}{\hbar}
\newcommand{\notp}{p \kern-.48em /}
\newcommand{\ci}{\cite}

\newcommand{\ot}{\otimes}

\newcommand{\bea}{\begin{eqnarray}}
\newcommand{\eea}{\end{eqnarray}}
\begin{document}
\setlength{\baselineskip}{1.5\baselineskip}
\thispagestyle{empty}
\title{Classical and quantum representation theory} \author{
N.P.~Landsman\thanks{ Supported by
an S.E.R.C. Advanced Research Fellowship}\\ \mbox{}\hfill \\
Department of Applied Mathematics and Theoretical Physics\\
University of Cambridge, Silver Street\\
Cambridge CB3 9EW, United Kingdom}
\date{}
\maketitle
\begin{abstract}
These notes present an introduction to an analytic version of deformation
quantization.
The central point is to study algebras of physical observables and their
irreducible representations.
In classical mechanics one deals with real Poisson algebras, whereas in quantum
mechanics the
observables have the structure of a real non-associative Jordan-Lie algebra.
The non-associativity
is proportional to $\pl^2$, hence for $\pl\raw 0$ one recovers a real Poisson
algebra. This
observation lies at the basis of `strict' deformation quantization, where one
deforms a given Poisson
algebra into a $C^*$-algebra, in such a way that the basic algebraic structures
are preserved.

Our main interest lies in degenerate Poisson algebras and their quantization by
non-simple
Jordan-Lie algebras. The traditional symplectic manifolds of classical
mechanics, and their quantum
counterparts (Hilbert spaces and operator algebras which act irreducibly)
emerge  from a generalized
 representation theory. This two-step procedure sheds considerable light on the
subject.

We discuss a large class of examples, in which the Poisson algebra canonically
associated to an
(integrable) Lie algebroid is deformed into the Jordan-Lie algebra of the
corresponding Lie groupoid.
A special case of this construction, which involves the gauge groupoid of a
principal fibre bundle,
describes the classical and quantum mechanics of a particle moving in an
external gravitational and
Yang-Mills field.
 \end{abstract}
 \newpage
\section{Introduction}
In quantization theory one tries to establish a correspondence between a
classical mechanical
system, and a quantum one. The traditional method, already contained in the
work of Heisenberg and
Dirac, is canonical quantization. Attempts to generalize  this procedure, and
put it on a solid
mathematical footing have led to geometric quantization \ci{Woo,Hur,GS}. This
is a certain algorithm
which still contains many gaps, and for various reasons cannot be considered
satisfactory \ci{Tuy}.
The same comment applies to path integral quantization, but we hasten to remark
that both
techniques have led to many examples, constructions,  and insights, in physics
as well as mathematics,
that would have been hard to reach otherwise, and still provide the main
testing ground for
alternative methods.

One such alternative method is deformation quantization.  The  version that we
use (and partly
propose)  employs techniques from algebra, differential geometry, and
functional analysis, and
appears to be very interesting from a mathematical point of view. One attempts
to relate Poisson
algebras to $C^*$-algebras in a way specified below, and as such it is possible
to relate to, and
exploit the phenomenal progress made in both subjects over the last decade.
This progress has consisted of  discovering and understanding general
structures through specific
examples, and in a certain sense a unification of the three mathematical
disciplines mentioned above
has been achieved, under the name of non-commutative geometry. On the
operator-algebraic side
this includes cyclic cohomology of operator algebras \ci{Con1} and operator
K-theory
(non-commutative topology) \ci{Ska}, which have found interesting applications
(highly relevant to
quantization theory!) in foliation theory and generalized index theorems
\ci{MS}. As to Poisson
algebras, we mention Poisson cohomology \ci{Hue} and the theory of symplectic
groupoids \ci{WX}.

{}From the point of view of physics we wish to stress that the quantization
procedure discussed here
is very satisfactory in that it
places physical notions like observables and states at the forefront (inspired
by algebraic quantum
field theory \ci{Haa}), plays down the (quite unnecessary) use of complex
numbers in quantum
mechanics, and accurately describes a large class of examples relevant to
Nature. Moreover, it brings
classical and quantum mechanics very closely together and highlights their
common structures.

We will introduce the relevant mathematical structures step by step, on the
basis of the familiar
Weyl quantization of a particle moving on $\Rn$. This will lead us to  Poisson
algebras, Jordan-Lie
algebras, and $C^*$-algebras. We then introduce the appropriate notion of a
representation of each
of these objects, and motivate an irreducibility condition. Lie groups form a
rich class of examples
on which to illustrate the general theory, but since these only describe
particles with nothing but
an internal degree of freedom, we must look elsewhere for structures describing
genuine physics.
A rich structure that is tractable by our methods, and at the same time
describes real physical
systems, is that of a Lie groupoid \ci{Mac,CDW}. It has an associated
`infinitesimal' object
(a Lie algebroid), and, as we will explain, the passage from an algebroid to a
groupoid  essentially
amounts to quantization.
\section{Classical mechanics and Poisson algebras}
\subsection{Introductory example: particle on flat space}
Consider a particle moving on the configuration space $Q=\Rn$. We use canonical
co-ordinates
$(\xm,\pmu)$ (usually simply written as $(x,p)$) on the cotangent bundle
$M=T^*\Rn$
($\mu=1,\ldots,n$), so that $(x,p)$ stands for the one-form $\pmu d\xm\in T^*_x
\Rn$. In mechanics a
key role is played by the Poisson bracket
\be
\{f,g\}=\frac{\na f}{\na \xm}\frac{\na g}{\na \pmu}-\frac{\na f}{\na
\pmu}\frac{\na g}{\na \xm}
, \ll{2.1}
\ee
where $f_1,f_2\in \cin(M)$. Here $\cin(M)\equiv A_0$ stands for the real vector
space of real-valued
smooth functions on $M$. Its elements are classical observables. Apart from the
Poisson bracket,
there is another bilinear map from $A_0\ot_{{\Bbb R}} A_0\raw A_0$, namely the
ordinary (pointwise)
multiplication $\cdot$. Let us write $f\sg g$ for $fg\, (\equiv f\cdot g)$, and
$f\al g$ for $\{f,g\}$.
The algebraic operations $\sg$ and $\al$ satisfy the following properties:
\begin{enumerate}
\item $f\sg g=g\sg f$ (symmetry);
\item $f\al g=-g\al f$ (anti-symmetry);
\item $(f\al g)\al h+(h\al f)\al g+(g\al h)\al f=0$ (Jacobi identity);
\item $(f\sg g)\al h=f\sg (g\al h)+ g\sg(f\al h)$ (Leibniz rule);
\item $(f\sg g)\sg h=f\sg(g\sg h)$ (associativity).
\end{enumerate}
The meaning of $\al$ and $\sg$ is as follows. To start with the latter, we
remark that the spectrum
${\rm spec}(f)$ of a function $f\in\cin(M)$ is the set of values it takes (that
is, the possible
values that the observable $f$ may have). If $f$ is concretely given (i.e., we
know ``$f(m_1)=a_1,\,
f(m_2)=a_2\ldots$'' then we obviously know the spectrum immediately. However,
$f$ may be regarded as
an abstract element of the algebra $A_0$. The point is now that ${\rm spec}(f)$
is completely
determined by its location in $A_0$, equipped with the product $\sg$
(forgetting the Poisson bracket).
Namely, if $a\in {\rm spec}(f)$ then $f-a1$ (where 1 is the function on $M$
which is identically
equal to 1) fails to have an inverse in $A_0$, whereas, conversely,
$(f-a1)^{-1}$ is a well-defined
element of $A_0$, satisfying  $(f-a1)^{-1}\sg (f-a1)=1$ if $a\not\in\spe(f)$.
Hence we may {\em
define} $\spe(f)$ as the set of real numbers $a$ for which $f-a1$ fails to have
an inverse in $A_0$.
A closely related point is that $\sg$ allows one to define functions of
observables (starting from
$f^2=f\sg f$); this is related to the previous point via the spectral calculus.

The Poisson bracket $\al$ determines the role any observable plays as the
generator of a flow on the
 space $M$ of pure states on $A_0$. To explain this, we need to introduce the
concept of the state
space of an algebra. The state space ${\cal S}(A)$ of a real  algebra $A$ may
be defined as the space
of normalized positive functionals on $A$, i.e., the linear maps $\om:A\raw
{\Bbb R}$ which satisfy
$\om(f^2)\geq 0$ for all $f\in A$, and $\om(1)=1$. If $\om_1$ and $\om_2$ are
states then $\lm
\om_1+(1-\lm)\om_2$ is a state if $\lm\in[0,1]$. A state is defined to be pure
if it does not allow
such a decomposition unless $\lm=0,1$; otherwise, it is called mixed. The
physical interpretation of
$\om(f)$ is that this number equals the expectation value of the observable $f$
in the state $\om$.
Any point $m$ of $M$ defines a pure state on $A_0$ by $m(f)=f(m)$, and these in
fact exhaust the set
of pure states. This statement holds equally well if we had taken $A_0$ to be
$\cci(M)$ or
$\cin_0(M)$ (the smooth functions with compact support, and those vanishing at
infinity,
respectively), but the pure state space of $\cin_b(M)$ (the bounded smooth
functions) is the
so-called Cech-Stone compactification of $M$.

Back to the Poisson bracket, each $f\in A_0$ defines a so-called Hamiltonian
vector field $X_f$ on $M$
by \be
X_fg=\{g,f\},\ll{2.2}
\ee
and this generates a Hamiltonian flow $\ph^f_t$ on $M$ (as the solution of the
differential equation
$d\phv^f_t/dt=X_f(\phv^f_t)$), cf.\ \ci{AM,LM}. That $X_f$ is indeed a vector
field (i.e., a
derivation of $\cin(M)$) is a consequence of the Jacobi identity on $\al$.

 The example $M=T^*\Rn$ has
the following feature: any two points of $M$ can be connected by a
(piecewisely) smooth Hamiltonian
flow. This property is equivalent to the following: $\{X_f(m)|f\in A_0\}=T_mM$
for all $m\in M$.
That is, the Hamiltonian vector fields span the tangent space at any point of
$M$.

To sum up, observables take values, and one may define functions of them, which
two properties are
determined by the product $\sg$; moreover,  they generate flows of the pure
state space, which are
determined by the Poisson bracket $\al$.
\subsection{Poisson algebras and their representations}
\begin{defn}
A Poisson algebra is a vector space $A$ over the real numbers, equipped with
two bilinear maps
$\al,\, \sg: A\ot_{{\Bbb R}}A\raw A$ which satisfy the five conditions in the
preceding subsection.
\end{defn}
The  examples of Poisson algebras we will consider are of the type $A=\cin(M)$
for some manifold
$M$, which has a Poisson structure, in the sense that $\al$ is some Poisson
bracket and $\sg$ is
multiplication. In that case, $M$ together with the Poisson structure is called
a Poisson manifold.
If $M$ has the special feature discussed after (\ref{2.2}) that any two points
can be joined by a
piecewisely smooth Hamiltonian curve, then $M$ is called symplectic. If not, we
can impose an
equivalence relation \ci{Wei83} $\sim$ on $M$, under which $x\sim y$ iff $x$
and $y$
can be joined by a piecewisely smooth Hamiltonian curve. The equivalence class
$L_x$ of any point can
be shown to be a manifold, which is embedded in $M$. If $i$ is the embedding
map then the relation
$\{i^*f,i^*g\}_{L_x}=i^*\{f,g\}_M$ defines a Poisson structure
$\{\;,\;\}_{L_x}$ on $L_x$, which is
obviously symplectic, and we call $L_x$ a symplectic leaf of $M$. More advanced
considerations
show that any Poisson manifold is foliated by its symplectic leaves \ci{LM}.

If $M=S$ is symplectic then the Poisson bracket can be derived from a
symplectic form on $S$
\ci{AM,LM}.
The corresponding  $A=\cin(S)$ are in some sense the `canonical models' of
Poisson algebras.
This motivates the following
\begin{defn}
A representation of a Poisson algebra $A$ is a map $\pi^S_c:A\raw \cin(S)$,
where $S$ is a
symplectic manifold, satisfying the following conditions for all $f,g\in A$:
\begin{enumerate}
\item $\pi^S_c(\lm f +\mu g)=\lm\pi^S_c(f)+\mu\pi^S_c(g)$, for all $
\lm,\mu\in{\Bbb R}$
(linearity);
\item $\pi^S_c(fg)=\pi^S_c(f)\pi^S_c(g)$ (preserves $\sg$);
\item $\pi^S_c(\{f,g\}_M)= \{\pi^S_c(f),\pi^S_c(g)\}_S$ (preserves $\al$);
\item The vector field $X_{\pi^S_c(f)}$ is complete if $X_f$ is
(self-adjointness).
\end{enumerate}
\ll{crep}
\end{defn}
The $c$ in $\pi_c^S$ stands for `classical', and the above defines a
`classical' representation
(as opposed to a `quantum' representation of algebraic objects by operators on
a Hilbert space; as
we shall see later, the distinction between classical and quantum is actually
blurred).
 A vector field
is called complete if its flow exists for all times. If $f$ had compact support
then its flow is
automatically complete \ci{AM}. Condition 4 excludes situations of the
following type.
Take $M=T^*{\Bbb R}$ with the usual Poisson structure (\ref{2.1}), and take $S$
any open set in $M$.
If $i$ is the embedding of $S$ into $M$, with the Poisson structure borrowed
from $M$ by restriction,
then $\pi_c^S(f)=i^*f$ satisfies 1-3 but not 4 (unless $S=M$).

The following theorem shows that all representations are actually of the type
$\pi_c^S=J^*$, where
$J:S\raw M$ is a Poisson morphism.
\begin{theorem}
Let $M$ be a finite-dimensional Poisson manifold, $A=\cin(M)$ the corresponding
Poisson algebra, and
let  $\pi^S_c:A\raw \cin(S)$ be a representation of $A$. Then there exists a
map $J:S\raw M$ such
that  $\pi_c^S=J^*$.
\ll{mom}
 \end{theorem}
{\em Proof.}
For the elementary  \ca theory used in the proof, cf.\ e.g.\ \ci{Ped,Tak,BR1}.
Take $s\in S$, and define a linear functional $\til{J}(s)$ on   $\cin_0(M)$ by
putting
$<\til{J}(s),f>=(\pi_c^S(f))(s)$ for $f\in\cin_0(M)$. By property 2 of a
representation,
$\til{J}(s)$ is multiplicative, hence positive and continuous, so it extends to
a pure state on the
commutative $C^*$-algebra $C_0(M)$ (which is the complexification of the
norm-closure of
$\cin_0(M)$).  Hence by the Gel'fand isomorphism $\til{J}(s)$ corresponds to a
point $J(s)$ of $M$.
Hence we have found the required map $J:S\raw M$. \Box

 For reasons to emerge in subsect.\ 2.3 below,  we will refer to $J$  as the
{\em generalized
moment map}. Property 3 of a representation implies that $J$ is what is called
a Poisson morphism.
Such maps have been studied extensively in the literature \ci{Wei83,LM}. The
self-adjointness
condition 4 translates into a condition on $J$, which is called {\em
completeness} by A. Weinstein.
Examples suggest that it is actually a classical analogue of the condition on
representations of
real operator algebras on Hilbert spaces that these preserve self-adjointness
(a special case of which
is the familiar requirement that group representations be unitary). However,
this self-adjointness
condition is actually a completeness condition, too, for it guarantees that the
unitary flow on
Hilbert space generated by the self-adjoint representative of a given operator
can be defined
for all times (also cf. sect.\ 3 below).
Further conditions on $\pi_c^S$ could be imposed to guarantee that $J$ is
smooth, but as
far as we can see we can develop the theory without those.

The following proposition (which is well known, cf.\ \ci{Wei83,LM}) is crucial
for the analysis of
irreducible representations (to be defined shortly). Here $J_*$ denotes the
push-forward of $J$
\ci{AM}.
 \begin{prop}
Let $J:S\raw M$ be the Poisson morphism corresponding to a representation
$\pi^S_c$ of the
Poisson algebra $\cin(M)$. Then for any $f\in\cin(M)$
\be
J_*X_{\pi_c^S(f)}=X_f, \ll{2vec}
\ee
where $X_f$ is the Hamiltonian vector field defined by $f$ (etc.).
Moreover, the image of the flow of $X_{\pi_c^S(f)}$ under $J$ is the flow of
$X_f$. \ll{flow}
\end{prop}
{\em Proof.}
Take $g\in\cin(M)$ arbitrary. By definition of $\pcs$ and $J$, we have
\be
\{\pcs(f),\pcs(g)\}_S(s)=\{f,g\}_M(J(s)) \ll{2.4}
\ee
Upon use of (\ref{2.2}), this leads to the identity
$(J_*X_{\pi_c^S(f)}g)(J(s))=(X_fg)(J(s))$, whence
the result. \Box

Since $S$ is symplectic, the symplectic form $\om$ provides an isomorphism
$\til{\om}:T^*_sS\raw
T_sS$ for any $s\in S$. This is given by $\til{\om}(df)=X_f$, or
$\til{\om}^{-1}(X)=i_X\om$
(evaluated at any point $s$). Now let $\til{T}_sS$ denote the subspace of
$T_sS$ which is spanned
by Hamiltonian vectors (i.e., of the form $X_{\pcs(f)},\, f\in\cin(M)$, taken
at $s$). Then
\be
\til{T}_sS=\til{\om}\circ J^*(T^*_{J(s)}M), \ll{2.5}
\ee
and $\til{\om}$ is a bijection between $\til{T}_sS$ and $ J^*(T^*_{J(s)}M)$,
where $J^*$ is the pull-back of $J$ (to 1-forms, in this case). This follows
rapidly from the
preceding proposition.

\begin{defn}
A representation $\pcs$ of a Poisson algebra $\cin(M)$ is called irreducible if
\ll{irc}
\end{defn}
\be
\{X_{\pcs(f)}(s)|f\in\cin(M)\}=T_sS\:\:\forall s\in S. \ll{2irr}
\ee
As mentioned before in a different variant, this condition guarantees that any
two points in $S$ can
be joined by a piecewisely smooth curve, whose tangent vector field is of the
form $X_{\pcs(f)}$.
Of course, since $S$ is symplectic any two points can be joined by such a curve
with tangent vectors
$X_g$, $g\in\cin(S)$, even if $\pcs$ is not irreducible, but one may not be
able to take $g=\pcs(f)$.
Note, that we could have broadened our definition of a representation by
allowing $S$ to be a
Poisson manifold; in that case, however, the irreducibility condition would
force $S$ to be
symplectic anyway. In the literature \ci{Wei83,LM} people appear to be mainly
interested in the
opposite situation, where a Poisson morphism $J: S\raw M$ ($S$ symplectic) is
called {\em full} if
(in our language) the corresponding representation $\pcs=J^*$ is faithful. As
the following result
shows, this is indeed quite opposite to an irreducible representation, which
has a large kernel
unless $M$ is  symplectic itself.
\begin{theorem}
If a representation $\pcs:\cin(M)\raw \cin(S)$ of a Poisson algebra is
irreducible then $S$ is
symplectomorphic to a covering space of a symplectic leaf of $M$.
\ll{leaf}
\end{theorem}
{\em Proof.}
We first show that $J:S\raw M$ is an immersion. Namely, if $J_*X=0$ for some
$X\in T_sS$ then
$ \langle J_*X,\th\rangle_{J(s)}=\langle X,J^*\th\rangle_s=0$, but by
(\ref{2.5}) and (\ref{2irr})
any $\th'\in T^*S$ may be written as $\th'=J^*\th$ for some $\th\in
T^*_{J(s)}M$. Hence $X=0$, and
$J$ is an immersion. Since $J$ is a Poisson morphism, it follows that $S$ is
locally symplectomorphic
to $J(S)\subset M$.

Next, $J(S)$ must actually be a symplectic leaf of $M$. For suppose that there
is a proper inclusion
$J(S)\subset L$, where $L$ is a symplectic leaf of $M$. It follows from the
Darboux-Weinstein theorem
\ci{AM,LM} that any point $x$ in a symplectic space has a neighbourhood $U_x$
such that any two points
in $U$ may be connected by a smooth Hamiltonian curve.If we take $x$ to lie on
the boundary of
$J(S)$ in $L$, then we find that there exist $m_1\in J(S)$ and $J(S) \not\ni
m_2 \in L$ which can be
connected by a smooth curve $\gm$  with tangent vector field $X_f$, for some
$f\in\cin(M)$. Let
$m_1=J(s_1)$, and consider the flow $\til{\gm}$ of $X_{\pcs(f)}$ starting at
$s_1$. By the
proposition above, $J\circ\til{\gm}=\gm$. However, since $m_2\not\in J(S)$, the
flow $\til{\gm}$ must
suddenly stop, which contradicts the self-adjointness  (completeness) property
4 of a representation.
Hence to avoid a contradiction we must have $J(S)=L$.

A similar argument shows that $J:S\raw J(S)$ must be a covering projection. For
$J$ not to be a
covering projection, there must exist a point $m\in M$, a neighbourhood $V_m$
of $m$, and a connected
component $J^{-1}_i(V_m)$ of  $J^{-1}(V_m)$, so that $J(J^{-1}_i(V_m))\subset
V_m$ is a proper
inclusion.
But in that case we could choose points $s_1\in J(J^{-1}_i(V_m))$ and
$J(J^{-1}_i(V_m))\not\ni
s_2\in V_m$ which can be connected by a smooth hamiltonian curve, and arrive at
a contradiction  to
the self-adjointness property of $\pcs$. \Box
\subsection{The Lie-Kirillov-Kostant-Souriau Poisson structure}
We obtain a basis class of Poisson algebras by taking $M=\g^*$, which is the
dual of the Lie algebra
$\g$ of some Lie group $G$. We may regard $X\in \g$ as an element of
$\cin(\g^*)$, by
$X(\th)=\langle\th,X\rangle$, and the Poisson structure of $\g^*$ is completely
determined by putting
\be
\{X,Y\}=[X,Y] \ll{2.0}
\ee
 (cf.\ \ci{AM,LM} for more information). The classical algebra of observables
$\cin(\g^*)$ describes
a particle which doesn't move, but only has an internal degree of freedom
(e.g., spin if $G=SU(2)$).

Let $\pcs:\cin(\g^*)\raw \cin(S)$ be a representation of $\cin(M)$, with $S$
connected. For each
$X\in\g$ we define a function $f_X$ on $S$ by
\be
f_X=\pcs(X). \ll{2.f}
\ee
By definition of a representation
\be
\{f_X,f_Y\}_S=f_{[X,Y]}. \ll{2.g}
\ee
If $\til{X}$ is the Hamiltonian vector field defined by $f_X$ (so that
$\til{X}g=\{g,f_X\}_S$) then
(\ref{2.g}) and the Jacobi identity imply that $[\til{X},\til{Y}]=-
\widetilde{[X,Y]}$ (where the
first bracket is the commutator of vector fields and the second one is the Lie
bracket on $\g$).
By self-adjointness,  the flow $\phv_t^X$ of $\til{X}$ is defined for all $t$,
and this leads to an
action  $\til{\pcs}$  of $\exp X\in G$ on $S$ by $\til{\pcs}(\exp
X)s=\phv_{1}^X(s)$. If $G$ is
simply connected this eventually defines a proper symplectic action of $G$ on
$S$.

Conversely, let $G$ act on a symplectic manifold $S$ so as to preserve the
symplectic form $\om$.
We may then define a vector field $\til{X}$ for each $X\in \g$ by
\be
(\til{X}f)(s)=\frac{d}{dt}f(e^{tX}s)_{|t=0}, \ll{2.h}
\ee
where we have written the action of $x\in G$ on $s\in S$ simply as $xs$. The
action is called
Hamiltonian  $i_{\til{X}}\om=df_X$ for some $f_X\in\cin(S)$ (this is guaranteed
if $H^1(S,{\Bbb
R})=0$), and strongly Hamiltonian if (\ref{2.g}) is satisfied  on top of that.
If the former condition
is met, one can define a map $J:S\raw \g^*$ by means of
\be
\langle J(s),X\rangle =f_X(s),\ll{2.i}
\ee
with pull-back $J^*:\cin(\g^*)\raw\cin(S)$.
In that case we clearly see from (\ref{2.f}) that the map $J$ defined
by (\ref{2.i}) is a special case of the generalized moment map constructed in
Theorem 1. Indeed, $J$
in (\ref{2.i}) is called the moment(um) map in the literature \ci{GS,AM,LM}.
(Note the varying sign
conventions. We follow \ci{AM} in putting $i_{\til{X}}\om =df_X$, $X_f
g=\{g,f\}$, and
$\{f_X,f_Y\}_S=f_{[X,Y]}$, but the alternative convention
 $i_{\til{X}}\om =-df_X$, $X_f  g=\{f,g\}$, and
$\{f_X,f_Y\}_S=-f_{[X,Y]}$ occurs as well.)

If the symplectic
$G$-action on $S$ is Hamiltonian but not strongly so, the right-hand side of
(\ref{2.g}) acquires an
extra term, and this situation may be analyzed in terms of Lie algebra
cohomology \ci{GS,AM,LM}. The
result is that the Poisson bracket (\ref{2.0}) can be modified, so that
$\pcs=J^*$ defines a
representation of $\cin(\g^*)$, equipped with the modified Poisson structure.

In the strongly
Hamiltonian case $J^*$ produces a representation $\pcs\equiv J^*$ of
$\cin(\g^*)$ equipped with the
Lie-Kirillov-Kostant-Souriau Poisson structure (\ref{2.0}). The fact that $J$
is a Poisson morphism
may be found in \ci{AM,LM,GS}, and it remains to check  the self-adjointness
condition.
We observe that vector fields on $S$ of the type $X_{\pcs(f)}$
($f\in\cin(\g^*)$) are tangent to a
$G$-orbit, so that their flow  $\gm_t^{\pcs(f)}$ cannot map a point of $S$ into
a different orbit.
This reduces the situation to the case where $G$ acts transitively on $S$. In
that case, the vector
fields $\{ \til{X}|X\in\g\}$ span the tangent space of $S$ at any point, so
that $\pcs$ is
irreducible.  By Theorem \ref{leaf}, the image of $J$ must be a symplectic leaf
of $\g^*$, hence a
co-adjoint orbit (this shows, incidentally, that the famous Kostant-Souriau
theorem which asserts
that any symplectic space which allows a transitive strongly Hamiltonian action
of a Lie group $G$
is symplectomorphic to a covering space of a co-adjoint orbit of $G$ \ci{GS,LM}
is a special case of
our  Theorem \ref{leaf}). Now take $f\in\cin(\g^*)$ with Hamiltonian vector
field $X_f$ and flow
$\gm_t^f$. Since (by definition) $G$ acts transitively on any co-adjoint orbit
in $\g^*$, we may
write $\gm_t^f(\th)=\pi_{\rm co}(x_t)\th$ for some curve $x_t$ in $G$ (not
uniquely defined, and
dependent on the argument $\th\in\g^*$); here
  $\pi_{\rm co}$ is the co-adjoint representation of $G$ on $\g^*$.
We now use Proposition \ref{flow} and the equivariance of $J$ (that is, $J\circ
x=\pi_{\rm co}(x)\circ J$ \ci{GS,AM,LM}) to derive $J\circ x_t(s)=J\circ
\gm_t^{\pcs(f)}(s)$ for any
$s\in S$; here $x_t$ depends on $J(s)$. Since $J$ is an immersion this implies
$\gm_t^{\pcs(f)}(s)=x_t(s)$, hence $\gm_t^{\pcs(f)}$ is defined whenever $x_t$
is; in particular, if
$\gm^f_t$ is complete  then $\gm_t^{\pcs(f)}$ is, so that the representation
$\pcs$ is self-adjoint.

In passing, we have observed that the irreducible representations of
$\cin(\g^*)$ are given by the
co-adjoint orbits in $\g^*$ (and their covering spaces).
 \section{Quantum mechanics and Jordan-Lie
algebras} \eo
\subsection{Weyl quantization on flat space}
To introduce some relevant mathematical structures in a familiar context we
briefly review the Weyl
quantization procedure of a particle moving on $Q=\Rn$, with phase space
$M=T^*\Rn$ (as in subsect.\
2.1). It is convenient to introduce a partial Fourier transform of
$f\in\cin(M)$ by
\be
\ff(x,\xd)=\int\frac{d^np}{(2\pi)^n}\, e^{ip\xd}f(x,p); \ll{3.1}
\ee
this makes $\ff$ a function on the tangent bundle $T\Rn$, where we use
canonical co-ordinates
$(x,\xd)\equiv \xd^{\mu}\partial/\partial \xm\in T_x\Rn$. For (\ref{3.1}) to
make sense we must have
that $f$ is integrable in the fiber direction (i.e., over $p$). Let $f$ be such
that $\ff\in
\cci(T\Rn)$; we refer to this class of functions as $\ovl{\A_0}$.  We then
define an operator
$\Qh(f)$ on the Hilbert space $\H=L^2(\Rn)$ by \be
(\Qh(f)\ps)(x)=\int d^ny\, \til{\Qh}(f)(x,y)\ps(y), \ll{3.2}
\ee
with kernel
\be
\til{\Qh}(f)(x,y)=\pl^{-n}\ff\left( \frac{x+y}{2},\frac{x-y}{\pl}\right).
\ll{3.3}
\ee
This operator is compact (it is even Hilbert-Schmidt, since the kernel in in
$\cci(\Rn\times \Rn$),
and thus it is bounded. (The norm of an operator $T$ on a Hilbert space $\H$ is
defined by $\n
T\n=\sup_{\ps} (T\ps,T\ps)^{1/2}$, where the supremum is over all vectors $\ps$
of unit length. An
operator $T$ is called bounded if this norm is finite. An operator is called
compact if it may be
approximated in norm by operators with a finite-dimensional range \ci{RS1}.
Compact operators
behave to some extent like finite-dimensional matrices).

A crucial property of $\Qh(f)$ is that it is self-adjoint (since $f$ is
real-valued). This means,
that $\Qh$ may be regarded as a map from $\ovl{\A_0}$ into
$\A=\K(L^2(\Rn))_{\rm sa}$ (the set of
self-adjoint compact operators on $\H=L^2$). As a real subspace of ${\cal
B}(\H)$, $\A$ is itself a
normed space, which is, in fact, complete (because $\K(\H)$ is). We can make
$\ovl{\A_0}$ into a
real Banach space, too, by equipping it with the norm
\be
\n f\n_0=\sup_{m\in M}|f(m)|. \ll{3.4}
\ee
The completion of  $\ovl{\A_0}$ under this norm is $\A_0=C_0(M,{\Bbb R})$ (the
space of real-valued
continuous functions on $M$ which vanish at infinity).

We interpret $\Qh(f)$ as the quantum observable corresponding to the classical
observable $f$.
Accordingly, we call $\A$ the (quantum) algebra of observables (of a particle
on $\Rn$). As in the
classical case, we may identify two algebraic operations on $\A$ (that is,
bilinear maps $\A\ot_{\Bbb
R}\A\raw \A$). They are
\be A\sh B= \half (AB+BA); \:\:\: A\ah B=\frac{1}{i\pl}(AB-BA). \ll{3.5}
\ee
The latter depends on $\pl$, so we will rename $\A$, equipped with $\sh$ and
$\ah$, as $\Ah$ (the
norm $\n \:\n$  does not depend on $\pl$). One may verify the following
properties:
\begin{enumerate}
\item $A\sh B=B\sh A$ (symmetry);
\item $A\ah B= -B\ah A$ (anti-symmetry);
\item $(A\ah B)\ah C+(C\ah A)\ah B+(B\ah C)\ah A=0$ (Jacobi identity);
\item $(A\sh B)\ah C=A\sh (B\ah C)+ B\sh(A\ah C)$ (Leibniz rule);
\item $(A\sh B)\sh C-A\sh(B\sh C)=\frac{\pl^2}{4}(A\ah C)\ah B$ (weak
associativity);
\item $\n A\sh B\n \leq \n A\n\,\n B\n$ (submultiplicativity of the norm);
\item  $\n A^2\n \leq \n A^2+B^2 \n$ (spectral property of the norm).
\end{enumerate}
We see that 1-4 are identical to the correpsponding properties of a Poisson
algebra, and 5 implies
that we are now dealing with a deformation of the latter in a non-associative
direction, in that the
symmetric product $\sh$ is now non-associative. A weak form of associativity
does hold, this is
 the so-called associator identity
  \be
(A^2\sh B)\sh A=A^2\sh (B\sh A), \ll{3.6}
\ee
which can be derived from 1-5. The last two properties imply $\n A^2\n=\n
A\n^2$ \ci{Upm}, which
leads to the usual spectral calculus.

Before commenting on the general structure we have found, let us find the
meaning of the products
$\sg_{\pl}$ and $\ah$ (cf.\ subsect.\ 2.1). We start with $\sh$. In classical
and quantum mechanics alike,
the spectrum of a self-adjoint operator is identified with the values the
corresponding observable
may assume. We have seen that the spectrum of a classical observable is
determined by the
symmetric product $\sg$. In standard Hilbert space theory (which is applicable,
as we have realized
$\A$ as a set of operators acting on $\H=L^2$) the spectrum  of a self-adjoint
operator $A$ is defined
as the set of values of $z$ for which the resolvent $(A-z)^{-1}$ fails to exist
as an element of
${\cal B}(\H)$ \ci{RS1}. More abstractly, the spectrum of an element $A$ of a
$C^*$-algebra ${\cal
B}$ is defined by replacing ${\cal B}(\H)$ by ${\cal B}$ in the above
\ci{Tak,BR1}. In fact, this
definition only uses the anti-commutator (rather than the associative operator
product, which
combines the anti-commutator and the commutator), so that we conclude that the
symmetric product on
the algebra of observables determines the spectral content. This observation is
originally due to
Segal \ci{Seg} (and was undoubtedly known to von Neumann, who introduced the
anti-commutator), and a
quick way to see this is that the spetcrum of $A$ is determined by the
$C^*$-algebra $C^*(A)$ it
generates; this is a {\em commutative} sub-algebra of $\cal B$ (or, $\K(\H)$ in
our example above)
which clearly only sees the anti-commutator $\sh$, which coincides with the
associative product on
$C^*(A)$ (cf. \ci[3.2]{SHH}). This argument is closely related to the fact that
the Jordan product
$\sh$ allows one to define functions of an observable, starting with $A^2\equiv
A\sh A$. Conversely,
one could start with a squaring operation, and define the Jordan product by
$A\sh
B=1/2((A+B)^2-A^2-B^2)$, cf.\ the Introduction of \ci{BR1}. The connection
between spectra and
functions of observables is provided by the spectral calculus.

Next, we wish to relate the commutator $\ah$ to the role observables play as
generators of
transformations of the space of pure states.  As explained prior to
(\ref{2.2}), we may introduce
states of an algebra of observables as normalized positive linear functionals
$\om$ on $\A$;
positivity here means that $\om(A^2)\geq 0$ for all $A\in\A$ (and $A^2=A\sh A$
as before), and
normalized means that $\n \om\n=1$ (which is equivalent to the property
$\om(1)=1$ if $\A$ has a
unit 1, which is not the case for $\A=\K(\H)$).
The state space of $\KH$ may be shown to be the space  of all denity matrices
on $\H$ (i.e., the
positive trace-class operators \ci{RS1} with unit trace). Pure states are as
defined before, and we
may consider the weak$\mbox{}^*$-closure $P(\A)$ of the set of all pure states
of $\A$. In our
example, any unit vector $\ps\in L^2$  defines a pure state $\om_{\ps}$ by
$\om_{\ps}(A)=(A\ps,\ps)$,
and, conversely, any pure state is obtained in this way. Noting that the space
of pure states thus
obtained is already weakly closed, we find that $P(\K(\H))$ is equal to  the
projective Hilbert space
$P\H$ (which by definition is the set of equivalence classes $[\ps]$ of vectors
of unit length, under
the equivalence relation $\ps_1\sim \ps_2$ if $\ps_1=\exp(i\al)\ps_2$ for some
$\al\in {\Bbb R}$). For
example, $P{\Bbb C}^2=S^2$ (the two-sphere)  is the pure state space of the
algebra of hermitian
$2\times 2$ matrices. More generally, $P\H$ is a Hilbert manifold modeled on
the orthoplement of an
arbitrary vector in $\H$. Hence $P {\Bbb C}^n$ is modeled on ${\Bbb C}^{n-1}$
To see this, take an
arbitrary vector $\chi\in\H$ (normalized to unity), and define a chart on the
open set $O_{\chi}\equiv
\{\ps\in\H\, |(\ps,\ch)\neq 0\}$ by putting $\Phi_{\ch}:O_{\ch}\raw
\chi^{\perp}$ equal to
$\Phi_{\ch}(\ps)=(\ps/(\ps,\ch))-\ch$. (We assume the inner product to be
linear in the first entry.)

The fundamental point is that $P\H$ has a Poisson structure \ci{AM}. To explain
this, note
first that $T_{\ps}\H\simeq \H$, since $\H$ is a linear space; a vector
$\phv\in\H$ determines
a tangent vector $\phv_{\ps}\in T_{\ps}\H$ by its action on any $f\in\cin(\H)$
\be
(\phv_{\ps}f)(\ps)=\frac{d}{dt}f(\ps+t\phv)_{|t=0}. \ll{ 3.7}
\ee
The symplectic form $\om$ on $\H$ is then defined by
\be
\om(\phv_{\ps},\phv_{\ps}')=-2\pl {\rm Im}\,(\phv,\phv').\ll{3.8a}
\ee
We now regard $A\in\A$ not as an operator on $\H$, but as a function
$\til{f}_A$ on $\H$, defined on
$\ps\neq 0$ by
\be
\til{f}_A(\ps)=\frac{(A\ps,\ps)}{(\ps,\ps)}. \ll{3.8}
\ee
(The value at $\ps=0$ is irrelevant). The point is that this definition
quotients to $P\H$, so that
$A\in\A$ defines a function $f_A$ on $P\H$ in the obvious way. Also, the
symplectic structure
quotients down to $P\H$ (the professional way of seeing this \ci{AM} is that
$U(1)$ acts on $\H$ by
$\ps\raw \exp(i\al)\ps$, this action is strongly Hamiltonian and leads to a
moment map $J: \H\raw
{\Bbb R}$ given by $J(\ps)=(\ps,\ps)$, and $P\H$ is the Marsden-Weinstein
reduction $J^{-1}(1)/U(1)$),
and this leads to the Poisson bracket
\be
\{f_A,f_B\}=f_{A\ah B}, \ll{3.9}
\ee
with $\ah$ defined in (\ref{3.5}). An analogous equation determines the Poisson
bracket on $\H$
itself. As explained in (\ref{2.2}) and below, the function $\til{f}_A$ (hence
$A$) defines a vector
field $\til{X}_A$ on $\H$, whose value at the point $\ps$ is found to be
 \be
\til{X}_A(\ps)=-\frac{i}{\pl}A\ps. \ll{3.10}
\ee
The flow $\til{\phv}^A_t$ of this vector field is clearly
\be
\til{\phv}^A_t(\ps)=e^{-itA/\pl}\ps. \ll{3.11}
\ee
Since this flow consists of unitary transformations of $\H$, it quotients to a
flow $\phv^A_t$ on
$P\H$, which is generated by a vector field $X_A$ which is just the projection
of $\til{X}_A$ to
the quotient space. This, in turn, is the vector field canonically related to
$f_A\in\cin(P\H)$ via
the Poisson structure (\ref{3.9}).

 Parallel to the discussion following (\ref{2.2}), we remark that
that $\A$ acts on $\H$ irreducibly, in the sense that any two points in (a
dense subset of) $\H$
may be connected by some flow generated by an element of $\A$. By projection, a
similar statement
holds for flows on $P(\A)=P\H$.  By (\ref{3.10}), this is equivalent to the
property that the
collection $\{A\ps|A\in\A\}$ is dense in $\H$ for each fixed $\ps$, and this,
in turn, by
(\ref{3.10}) is exactly the irreducibility condition used in Definition
\ref{irc} for Poisson
algebras.

To sum up, we have shown that the product $\ah$ indeed leads to the desired
connection between
observables and flows on the pure state space of $\Ah$ (note that all the $\Ah$
are isomorphic to
$\A$ for $\pl\neq 0$), just as in the classical case.

Remarkably, the Jordan product $\sh$ has a geometric expression in terms of the
functions $f_A$ on
$P\H$, too \ci{CMP}. Let $g$ be the K\"{a}hler metric on $\H$, which is defined
by (cf.\
(\ref{3.8a}))
\be
g(\phv_{\ps},\phv_{\ps}')=\pl {\rm Re}\,(\phv,\phv').\ll{kahler}
\ee
Then a calculation shows that
\be
f_{A\sh B}=\pl g(\til{X}_A,\til{X}_B)+f_A f_B; \ll{metric}
\ee
this should be compared with (\ref{3.9}), which for this purpose may be
rewritten as
\be
f_{A\ah B}=\om(\til{X}_A,\til{X}_B)+0. \ll{symp}
\ee
We see that the entire Jordan-Lie algebraic structure of $\A$ is encoded in the
K\"{a}hler structure
of $P\H$, which is given by hermitian metric $\Omega$ defined by the inner
product:
\be
\Omega (\phv_{\ps},\phv_{\ps}')= \pl(\phv,\phv').
\ee
Clearly, $\Omega=g-\half i\om$.
 \subsection{Jordan-Lie algebras}
 We now generalize some of  these considerations.
\begin{defn}
A real Banach space $\A$ equipped with two bilinear maps $\sh,\,\ah:$\\ $
\A\ot_{\Bbb
R}\A\raw \A$, which satisfy properties 1-7 in the preceding subsection, is
called a Jordan-Lie
algebra. If $\pl\neq 0$ $\A$ is called  non-associative, and if $\pl=0$ $\A$ is
called
associative. In the latter case the operation $\al_0$ is only required to be
densely defined.
\ll{JL}
\end{defn}
 The Jordan-Lie structure of von Neumann's choice of ${\cal B}(\H)$ as the
algebra of observables in
quantum mechanics was emphasized in \ci{Emc}. We here propose that Jordan-Lie
algebras are the
correct choice to take as algebras of observables in quantum mechanics;
allowing more possibilities
than  ${\cal B}(\H)_{\rm sa}$ or $\K(\H)_{\rm sa}$ allows the incorporation of
superselection rules,
and the quantization of systems on topologically nontrivial phase spaces
\ci{NPLI}.
The example above already illustrates the remarkable fact that {\em
conventional quantum mechanics may
be described without the use of complex numbers}.  The reader may object that a
factor $i$ appears
in (\ref{3.5}), but the resulting product $\ah$ maps two self-adjoint operators
into a self-adjoint
operator, and it is the algebraic structure on $\A$ (given by $\sh$ and $\ah$),
a real vector space,
which determines all physical properties. Also, the (pure) state space is a
real convex space and
all observable numbers in quantum mechanics are of the form $\om(A)$, where
$\om$ is a state and $A$
an observable.

A first major advantage of starting from Jordan-Lie algebras is that Poisson
algebras are a special
 case (in which the symmetric product is associative), obtained by putting
$\pl=0$ in property 5.
Hence classical and quantum mechanics are described by the same underlying
algebraic structure (of
which the former represents a limiting case), a point not at all obvious in the
usual description in
terms of either symplectic manifolds or Hilbert spaces.

A second comment is that the axioms imply that $\A$ must the self-adjoint part
of a $C^*$-algebra,
so that we recover a mathematical structure that has proved to be exceptionally
fruitful in the
study of quantum mechanics \ci{Seg,NPLI}, quantum field theory \ci{Haa},
statistical mechanics
\ci{BR2,Haa}, and pure mathematics \ci{Con1,Con82,MS,Ska,Tak,Ped}. Indeed, we
may define an
associative multiplication on $\A_{{\Bbb C}}=\A\ot_{{\Bbb R}}{\Bbb C}$ by means
of
\be
AB=A\sh B+\half i\pl A\ah B; \ll{3.12}
\ee
the associativity follows from the axioms, cf.\ \ci{Emc}.  The involution in
$\A_{{\Bbb C}}$ is
simply given by the extension of $A^*=A$  for $A\in\A$. The norm axioms imply
that  $\A_{{\Bbb C}}$
thus obtained is a $C^*$-algebra.

The meaning of $\sh$ and $\ah$ is the same as in the example of the compact
operators.  To explain
this, it is convenient to use `Kadison's function representation' \ci{Kad51} of
the self-adjoint part
of any  $C^*$-algebra (hence of any Jordan-Lie algebra). Let $K$ be the state
space of $\A$ (equipped
with the weak$\mbox{}^*$-topology); this space is compact if $\A$ has a unit,
which we shall assume
(if not, one can adjoin one in a canonical way without any loss of information
\ci{Tak,SHH}). Then
$\A$ is isometrically isomorphic with the space $A(K)$ of all affine
real-valued continuous functions
on $K$ (with norm given by the supremum); since $K$ is a convex subspace of the
linear space of all
continuous linear functionals on $\A$, convex combinations
$\lm(\om_1)+(1-\lm)\om_2$ ($\lm\in[0,1]$)
of states are well-defined, and a function $f$  on $K$ is called affine if
$f(\lm(\om_1)+(1-\lm)\om_2)=\lm f(\om_1)+(1-\lm)f(\om_2)$ for all $\om_i\in K$
and all $\lm\in[0,1]$
(cf.\ \ci[III.6]{Tak} for detailed information on such spaces).  The
isomorphism between $A\in\A$ and
$\til{A}\in A(K)$ is simply given by $\til{A}(\om)=\om(A)$.
The spectral theory of $\A$, which, as we have seen in the case $\A=\K(\H)_{\rm
sa}$, is governed by
the symmetric product $\sh$ (using an argument which extends to the general
case), translates into a
spectral theory for such affine functions \ci{AS}. Conversely, if one starts
from $A(K)$ as the
basic structure, one may set up a spectral calculus, which exploits the very
special properties that
$K$ has because it is the state space of a $C^*$-algebra (hence, in particular,
of a Jordan algebra).
This spectral theory may then be used to {\em define} $\sh$ \ci{AS}, making the
intimate connection
between the symmetric product and the spectral calculus even clearer than in
the realization of $\A$
as operators on a Hilbert space.

By the affine property, an element of $A(K)$ is completely determined by its
values on the pure
state space $P(\A)$ (which is the $w^*$-closure of the extreme boundary of $K$
\ci{Tak,Ped}). We can
define an equivalence relation $\sim$ on $P(\A)$, saying that $\om_1\sim\om_2$
if both states give
rise to unitarily equivalent representations (via the GNS construction, which
provides a connection
between states and representations \ci{BR1,Tak}). Each equivalence class
defines a so-called folium of
$P(\A)$.
Each such folium is a Hilbert manifold, which is diffeomorphic (hence affinely
isomorphic) to the
pure state space $P(\H)$ for some Hilbert space $\H$ (cf.\ the previous
subsection). Therefore,  it
admits a Poisson structure, which is defined exactly as in the case $\A=\K(\H)$
(the compactness of
$A$ and $B$ in (\ref{3.9}) was not essential). The Poisson structures on the
folia can be combined
into a Poisson structure on $P(\A)$, which is degenerate iff $\A$ (unlike the
compact operators)
admits more than one equivalence class of irreducible representations. This
eventually leads us to
regard elements of $A(K)$ (hence of $\A$) as generators of transformations of
$P(\A)$, and we see
that the flow of a given operator cannot leave a given folium. This suggests
that $P(\A)$ is a
Poisson manifold, which is foliated by the symplectic leaves $P(\H)$, but much
remains to be done
before this statement can be made precise, let alone proved (the main problems
are to patch the
folia together in the weak$\mbox{}^*$-topology on $P(\A)$, and to deal with the
states that are not
pure but are weak$\mbox{}^*$ limits of pure states. In the uniform topology on
$K$ and $P(\A)$ things
are easy, because $P(\A)$ splits up as a collection of disjoint components,
each component being a
folium, but this topology is not the relevant one).

Thus the idea is to identify the inequivalent irreducible representations of
$\A$ (that is, its
superselection sectors \ci{Haa}) with the symplectic leaves of the pure state
space $P(\A)$,
providing a nice parallel with the classical case. The total state space $K$ of
$\A$ may be equipped
with a Poisson structure, too, but it is clear that the symplectic leaves of
this Poisson space
cannot be identified with inequivalent representations. For already in the
simplest case where $\A$
consists of the hermitian $n\times n$ matrices the state space is foliated by
an uncountable number
of symplectic leaves, whereas the inequivalent representations are labeled by a
positive integer.
(To see this, note that $K$ can be embedded in the dual ${\bf u}(n)^*$ of the
Lie algebra of $U(n)$,
equipped with the canonical Lie-Poisson structure, and this embedding is a
Poisson morphism. Hence the
symplectic leaves of $K$ are simply given by those leaves of  ${\bf u}(n)^*$
which lie in $K$; these
are generalized flag manifolds, and there are uncountably many even of a given
orbit type).

In any case, we see that the role of the antisymmetric product $\ah$ as the
agent which relates
observables to flows on the pure state space survives unscratched for
Jordan-Lie algebras.
Conversely, we would like to define this product in terms of the Poisson
structure on $P(\A)$.
This can presumably be done using a result of Shultz \ci{Shu}, who proved that
the commutator on the
self adjoint part  $\A$ of   a $C^*$-algebra $\A$ is {\em abstractly}
determined by specifying
transition probabilities and an orientation on $P(\A)$. These transition
probabilities are the usual
ones if one passes from states to their GNS representations (and are zero for
disjoint states, that
is, states leading to inequivalent representations).   Specifying
$|(\ps_1,\ps_2)|^2$ plus an
orientation is equivalent to specifying ${\rm Im}\, (\ps_1,\ps_2)$, so we see
from (\ref{3.8a}) that
the theorem in \ci{Shu} can very simply be understood by saying that the
commutator is given by the
Poisson bracket (\ref{3.9}), and that Poisson and Jordan isomorphisms between
two state spaces are
induced by isomorphisms of the corresponding Jordan-Lie algebras.

 We return to the axioms 1-7 on a Jordan-Lie algebra. Especially the norm
axioms, but also property
5 look rather arbitrary, and it would be nice to reformulate them in such a
way, that the following
question may be answered: {\em which physical postulates  of quantum mechanics
imply its description
in terms of Jordan-Lie algebras and their state spaces?}. A similar question
concerned with the
Hilbert space formulation of conventional quantum mechanics is analyzed in
\ci{Lud,BC}.
Since a Jordan-Lie algebra is isomorphic to the self-adjoint part of a
$C^*$-algebra, we can look at
the literature for help. In turns out to be fruitful to shift emphasis from the
Jordan-Lie algebra
$\A$ to its state space $K$ (from which $\A$ can be recovered as $A(K)$, as we
have reviewed above).
The question above may then be reformulated by asking which properties of a
compact convex set $K$
make $A(K)$ isomorphic to a Jordan-Lie algebra, and what the physical meaning
of these properties is
(as before, we stress the point that by eliminating complex numbers and Hilbert
spaces from quantum
mechanics through its reformulation in terms of Jordan-Lie algebras and their
state spaces, we feel
that we have come closer to the physical meaning of this theory).

The latter question has partly been answered in the work of Alfsen and Shutz
\ci{AS,AS80,Shu}, and
others (cf.\ the reviews \ci{Alf,Upm}). As a consequence of these papers, the
origin of the Jordan
structure in quantum mechanics (as well as the norm axioms, which only use the
Jordan product $\sh$)
is now quite well understood. The key property of $K$ that leads to a Jordan
structure and the
associated spectral calculus is the existence of sufficiently many projective
faces in $K$; a
projective face plays a role similar to that of a closed subspace of a Hilbert
space (or the
corresponding projector) and is physically a yes-no question. Projective faces
are
orthocomplemented, and have other nice properties making them suitable as a
basic ingredient of
quantum logic \ci{BC,Coh}. Other properties of $K$ which are necessary to
derive the Jordan structure
are related to the property that pure states in quantum mechanics can be
prepared through filtering
procedures, and to the symmetry of transition amplitudes (which reflects the
symmetry between pure
states and finest detectors \ci{Haa}).

Further properties of the state space $K$ leading to a Lie bracket on $\A\simeq
A(K)$ are known
\ci{AS80,Alf}, but their physical meaning is not so clear. We hope to be able
to show that these
properties are equivalent to $P(\A)$ admitting a Poisson structure which
foliates the pure state
space in a way consistent with the representation theory of $\A$ as a Jordan
algebra.
A crucial property of non-associative Jordan-Lie algebras (i.e., $\pl\neq 0$)
is that the restriction
of $A(K)$ to $P(\A)$ does not coincide with the space of all continuous
functions on $P(\A)$
(unlike the classical case; the essential point is that not nearly every
function on $P(\A)$
extends to an affine function on $K$, because non-pure elements of $K$
generically have many
decompositions as convex sums of pure states \ci{BR1,Tak}. This non-uniqueness
constrains the
allowed functions on the extreme boundary $P(\A)$ of $K$, which do have an
affine extension to $K$,
enormously. Such constraints do not arise when every mixed state in $K$ has a
unique extremal
decomposition, and this happens precisely when $\A$ is associative, i.e., in
the classical case.).
Together with the Poisson structure this property should be related to the
uncertainty principle (at
least in its naive textbook formulation).
 \subsection{Representation theory of Jordan-Lie algebras}
As we have seen in Definition (\ref{crep}), a representation of a Poisson
algebra is a map into
$\cin(S)$ for some symplectic space $S$, which preserves  all the algebraic
structures and in
addition satisfies a completeness condition. The motivation was that $\cin(S)$
for symplectic $S$ is
a `canonical' model of a Poisson algebra. More importantly, irreducibility
implies that $S$ must be
symplectic. Similarly, the canonical model of a Jordan-Lie algebra is the
algebra of all bounded
self-adjoint operators on a complex Hilbert space $\H$. The latter is naturally
equipped with the
Jordan-Lie structure (\ref{3.5}), and this motivates
\begin{defn}
A representation of a non-associative Jordan-Lie algebra $\A$ is a  map
$\pi_q^{\ch}: \A\raw
{\cal B}(\Hs)$, for some Hilbert space $\Hs$, satisfying for all $A,\, B\in\A$
\begin{enumerate}
\item $\pqs(\lm A+\mu B)=\lm \pqs(A)+\mu\pqs(B)$ (linearity);
\item $\pqs(A\sh B)=\half(\pqs(A)\pqs(B)+\pqs(B)\pqs(A))$ (preserves Jordan
product);
\item $\pqs(A\ah B)=\frac{1}{i\pl}(\pqs(A)\pqs(B)-\pqs(B)\pqs(A))$ (preserves
Lie product);
\item $\pqs(A)^*=\pqs(A)$ (self-adjointness).
\end{enumerate}
\ll{qrep}
\end{defn}
These conditions are, of course, equivalent to the usual ones on
representations of the \ca
$\A_{{\Bbb C}}$ (the self-adjointness condition 4 is the requirement that
$\pqs$ is a
$\mbox{}^*$-representation of $\A_{{\Bbb C}}$), but we have put them in the
given form to make the
analogy with the classical Definition \ref{crep} clear.  In similar vein, the
classical
irreducibility condition Definition \ref{irc} is, as we have seen from the
discussion following
(\ref{3.11}), essentially the same as the usual definition of irreducibility
for representations of
$C^*$-algebras, which in the present framework reads \begin{defn}
A representation $\pqs$ of a Jordan-Lie algebra $\A$ on a Hilbert space $\Hs$
is called irreducible
iff every vector in $\Hs$ is cyclic for $\pqs(\A)$ (that is, the set
$\{A\ps|A\in\A\}$ is dense in
$\Hs$ for each fixed $\ps\in\Hs$).
\ll{irq}
\end{defn}
All this may be reformulated in terms of the (pure) state space of $\Hs$, and
the Jordan and Lie
products on $A(K)$ as discussed in the previous subsection, but we leave this
to the reader.

There is a decisive difference between the classical case ($\pl=0$; Jordan
product $\sg\equiv \sg_0$
associative) and the quantum case as far as irreducibility is concerned.
Irreducible representations
of a Poisson algebra $\cin(M)$ are highly reducible as representations of the
corresponding Jordan
algebra (in which the anti-symmetric product $\al$ is ignored), whereas
irreducible representations
of this Poisson algebra (which are just points of $M$) do not lead to
representations of $\cin(M)$
at all. In the quantum case, a representation of a  non-associative  Jordan-Lie
algebra is
irreducible iff it is irreducible as a representation of the underlying Jordan
algebra. This looks
curious, because the irreducibility condition  above may be formulated in terms
of the vector
fields (\ref{3.10}), which are defined using the Lie product (see (\ref{3.9})).
However, the
unitary flow (\ref{3.11}) is completely defined in terms of the Jordan product
(which allows the
definition of functions of an operator).

    The naive quantum analogue of the generalized moment map $J$
(cf.\ Theorem \ref{mom}) is rather trivial: given a representation $\pqs(\A)$,
we may define a map
$\til{J}:\Hs\raw K$ (where $K$ is the state space of $\A$) by specifying the
value of the state
$\til{J}(\ps)$ on arbitrary $A\in\A$ to be
 \be
(\til{J}(\ps))(A)=\frac{(\pqs(A)\ps,\ps)}{(\ps,\ps)}. \ll{3.mom}
\ee
This evidently reduces to a map $J:P\Hs\raw K$, which is the naive quantum
analogue of the
classical generalized moment map. Namely, for $\pqs$ irreducible, the image of
$J$ is contained in the
pure state space $P(\A)$.  Thus we see that the quantum moment map just
expresses the correspondence
between states and vectors in a Hilbert space, which  is central to the GNS
construction
\ci{Tak,BR1}, and lies at the heart of operator  algebras. A difference
beteween the classcial and
the naive quantum moment map is that the image of the former is the set of pure
states, even if the
representation is reducible, while the image of the latter may well lie among
the mixed states
(namely if the representation is reducible). Also, the Marsden-Weinstein
symplectic reduction
construction \ci{AM,LM} canot be `quantized' in terms of $\til{J}$ in any
obvious way.
Hence one needs a deeper quantum analogue of  the classical moment map, and
this is given by the
concept of a Hilbert $C^*$-module, see \ci{RMW}.

 The quantum
counterpart of the classical Theorem \ref{leaf} has yet to be proved (and even
properly formulated);
this would express that $P(\A)$ is foliated by its symplectic leaves, which, as
we have seen in the
preceding subsection, should be identified with folia of states leading to
equivalent representations.
\subsection{The group algebra} For reasons to emerge later, a quantum analogue
of the Poisson algebra
$\cin(\g^*)$ (cf.\ subsect.\ 2.3) is the group algebra $JL(G)=C^*(G)_{\rm sa}$;
it is the quantum
algebra of observables of a particle whose only degree of freedom is internal.
Here $G$ is any Lie
group with Lie algebra $\g$. For simplicity, we only define $JL(G)$ for
unimodular $G$ (look up
$C^*(G)$ in  \ci{Ped} for the general case). The starting point is to construct
a dense subalgebra of
$C^*(G)$.This is done by defining a product $*$ and involution $\mbox{}^*$ on
$\cci(G)$ by
\be
(f*g)(x)=\int_G dx\, f(xy)g(y^{-1}); \:\:\: f^*(x)=\ovl{f(x^{-1})},
 \ll{3G1}
\ee
where $dx$ is a Haar measure on $G$. The norm is defined in \ci{Ped}; in the
special case that $G$
is amenable (this holds, for example, when $G$ is compact) one may put $\n
f\n=\n \pi_q^L(f) \n$,
where $\pi_q^L$ is a representation of $\cci(G)$ (regarded as an associative
$\mbox{}^*$-algebra)
on
$\H_L=L^2(G)$, given by
 \be
(\pi_q^L(f)\ps)(x)=\int_G dy\, \ f(y) (\pi_L(y)\ps)(x), \ll{3G2}
\ee
with $ (\pi_L(y)\ps)(x)=\ps(y^{-1}x)$.
The closure of $\cci(G)$ in this norm is the group algebra $C^*(G)$. The
corresponding Jordan-Lie
algebra $JL(G)$ is its self-adjoint part, equipped with the products $\sh$ and
$\ah$, defined as in
(\ref{3.5}) (with $AB$ replaced by $f*g$, etc.).

The representation theory of $JL(G)$ coincides with that of $C^*(G)$, which is
well-known \ci{Ped}:
every (non-degenerate) representation $\pi_q^{\ch}$  of $JL(G)$ on a Hilbert
space $\Hs$ corresponds
to a unitary representation $\pi_{\ch}$ of $G$ on $\Hs$, the passage from
$\pi_{\ch}(G)$ to
$\pqs(JL(G))$ being accomplished by the analogue of (\ref{3G2}), with $L$
replaced by $\ch$.
In particular, irreducible representations of $JL(G)$ correspond to irreducible
unitary
representations of $G$.

In traditional quantization theory (applied to this special case) one tried to
associate a Hilbert
space and certain operators to a co-adjoint orbit $\O\subset\g^*$ and the
associated Poisson algebra
$\cin(\O)$ (which we look upon as an irreducible representation of
$\cin(\g^*)$). This was very
succesful in special situations, e.g., $G$ nilpotent. In that case there is a
one-to-one
correspondence between co-adjoint orbits and unitary representations, given by
the Dixmier-Kirillov
theory \ci{CG}. The same strategy was reasonably succesful in some other cases,
like $G$ compact and
semi-simple, when any irreducible unitary representation of $G$ can be brought
into correspondence
with at least some co-adjoint orbit via the Borel-Weil theory \ci{Hur}; on the
other hand, most
co-adjoint orbits do not correspond to any unitary  representation of $G$ at
all. However, in the
general case no correspondence between co-adjoint orbits and irreducible
representations exists, and
modern research in representation theory looks in different directions \ci{Vog}
(note that this does
not undermine the hard fact that the classical irreducible  representations of
$\cin(\g^*)$ are
completely classified by the co-adjoint orbits and their covering spaces).

 The natural correspondence between classical and quantum mechanics exists at
an
algebraic level, namely in their respective Jordan-Lie algebras of observables.
The irreducible
representations of a classical Poisson algebra are not necessarily related to
those of the
corresponding quantum Jordan-Lie algebra, and both should be constructed in
their own right.
\section{Quantization}
\eo
\subsection{The definition of a quantization}
We now return to the Weyl quantization on $\Rn$ reviewed in subsect.\ 3.1. We
have seen how we may
regard $\Qh$ as a map from the dense subspace $\ovl{\A_0}$ of the commutative
Banach algebra
$\A_0=C_0(T^*\Rn)$ to the space of self-adjoint compact operators
$\A=\K(L^2(\Rn))_{\rm sa}$.
Here $\A_0$ also has a densely defined Poisson structure (which is, in
particular, defined on
 $\ovl{\A_0}$), and may be regarded as an associative Jordan-Lie algebra,
equipped with the products
$\cdot=\sg\equiv \sg_0$ and $\{\;,\;\}=\al\equiv \al_0$. The space $\A$ may be
dressed up with the
products $\sh$ and $\ah$, defined in  (\ref{3.5}), and thus a family of
non-associative Jordan-Lie
algebras $\{\Ah\}$ is defined (the norm in $\Ah$ is borrowed from $\A$, and is
independent of
$\pl$ for $\pl\neq 0$. The norm on $\A_0$ is  defined in (\ref{3.4})).  We
define $Q_0: \A_0\raw
\A_0$ as the identity map. It may be shown \ci{NPLstr} that the following
properties hold for all
$f,g\in \ovl{\A_0}$:
 \begin{enumerate}
\item
$\lpl \n \Qh(f)\sh \Qh(g)-\Qh(f\sg_0 g)\n =0$;
\item $\lpl \n \Qh(f)\ah \Qh(g)-\Qh(f\al_0 g)\n =0$;
\item the function $\pl\raw \n \Qh(f) \n$ is continuous on $I={\Bbb R}$.
\end{enumerate}
Condition 2 is an analytic reformulation of the correspondence between
commutators of operators and
Poisson brackets of functions first noticed by Dirac. The first condition is
based on the
correspondence between anti-commutators of operators and pointwise products of
functions, first
noticed by von Neumann. The third condition is a precise formulation of (one
form) of the
correspondence principle due to Bohr.
Recalling that $f\sg_0 g=fg$ and $f\al_0 g=\{f,g\}$, note the consistency of
the above conditions
with (\ref{metric}) and (\ref{3.9}).  In the context of $C^*$-algebras
conditions 2 and 3 in
their present form were first written down by Rieffel \ci{Rie1} (who did not
impose either condition
1 or self-adjointness on a quantization map). The connection between
deformations of algebras and
quantization theory was analyzed in a different mathematical setting in
\ci{Ber,Bay}.

The example of a particle on $\Rn$ and the general considerations in sections 2
and 3 motivate the
following
\begin{defn}
Let $\A_0$ be a commutative Jordan algebra with a densely defined Poisson
bracket (making $\A_0$
into an associative Jordan-Lie algebra, cf.\ Def.\ \ref{JL}), and let
$\ovl{\A_0}$ be a dense
subalgebra on which the Poisson bracket is defined. A quantization of this
structure is a family
$\{\Ah\}_{\pl\in I}$ of  non-associative Jordan-Lie algebras (Def.\ \ref{JL}),
and a family $\{\Qh
\}_{\pl\in I}$ of maps defined on $\ovl{\A_0}$, such that the image of $\Qh$ is
in $\Ah$, and the
above conditions 1-3 are satisfied.
\ll{qua}
\end{defn}
As we have seen, Weyl quantization satisfies this definition. A generalization
of Weyl quantization
to arbitrary Riemannian manifolds is given in \ci{NPLstr}. The axioms above are
not quite satisfied
by this generalized quantization prescription, in that the range in $\pl$ for
which $\Qh$ is defined
depends on its argument. This is easily remedied, however, by constructing
cutoff functions in
$\pl$, cf.\ the example below. The cutoff, on the other hand, upsets the
physical interpretation of
$\Qh(f)$ as the quantum observable corresponding to the classical observable
$f$ for all $\pl\in I$,
and for that reason in \ci{NPLstr} we preferred to leave $\Qh(f)$ undefined
whenever it could no
longer by interpreted properly. This complication only occurs for manifolds for
which the exponential
map is not a diffeomorphism on the entire tangent space at each point. A
further generalization is
to admit internal degrees of freedom, through which the particle can couple to
a gauge field. This
case is covered in \ci{NPLstr}, too, and from this general class of examples it
has become clear that
the definition of quantization given above is satisfied by a number of
realistic physical examples.

A non-self-adjoint version of the quantization of $C_0(\g^*)$ by $C^*(G)$ (cf.\
subsects.\  2.3 and
3.4) was first given by Rieffel \ci{Rie2}, and the physically relevant
self-adjoint version, i.e.,
the construction of the maps $\Qh: C_0(\g^*)\raw JL(G)$ is a special case of
the theory in
\ci{NPLstr} if $G$ is compact (obtained by taking $P=H=G$ in that paper, and
exploiting the fact
that $(T^*G)/G\simeq \g^*$ with the usual Poisson structure). We define
$\ovl{\A_0}\subset \til{C_0}(\g^*)$  as the space of those functions $f$  on
$\g^*$ whose Fourier
transform $\ff$ is in $\cci(\g)$ (since $\g^*\simeq \Rn$ we can define the
Fourier transform as
usual, cf.\ (\ref{3.1}), omitting the $x$-dependence). The quantization map is
given by
\be
(\Qh(f))(e^{-\pl X})=\pl^{-n}\ff(X), \ll{g1}
\ee
which defines the left-hand side as an element of $C^*(G)_{\rm sa}= JL(G)$ for
those values
of $\pl$ for which $\pl$ times the support of $\ff$ lies in the neighbourhood
of $0\in\g$ on which
the exponential function is a diffeomorphism from $\g$ to $G$. Since $\ff$ has
compact support, the
allowed values of $\pl$ will lie in an interval $(-\pl_0,\pl_0)$, where $\pl_0$
depends on $f$.
If the group $G$ is exponential (which is the case if $G$ is simply connected
and nilpotent \ci{CG})
then $\pl_0=\infty$.  In general, one could extend the quantization to any
value of $\pl$, without
violating the conditions required by Def.\ \ref{qua}, by multiplying $\Qh(f)$
by a function $h$
which is 1 in  $(-.99 \pl_0,.99 \pl_0)$ (say).
\subsection{Positivity and continuity}
 While the Weyl quantization of subsect.\ 3.1 (as well as its generalization to
Riemannian manifolds)
satisfies Def.\ \ref{qua} of a quantization, there are two serious problems
with it. The first is
lack of positivity; this means that if $f\geq 0$ in $\A_0 =C_0(T^*\Rn)$ then it
is not necessarily
true that $\Qh(f)\geq 0$ in $\A$ (see e.g. \ci[2.6]{Fol}). From the equality
\be
(\Qh(f)\Om,\Om)=\int_{T^*\Rn} \frac{d^nx d^np}{(2\pi)^n}\,
\wi_{\Om}(x,p)f(x,p),\ll{p1}
\ee
with the Wigner function
\be
\wi_{\Om}(x,p)=\int_{\Rn} d^n \xd\,
e^{ip\xd}\Om(x-\half\pl\xd)\ovl{\Om(x+\half\pl\xd)}, \ll{p2}
\ee
we see that the potential non-positivity of $\Qh(f)$ is equivalent to the fact
that the Wigner
distribution function (\ref{p2})  is not necessarily  positive definite.

The second problem is that $\Qh$ (for fixed $\pl\neq 0$) is not continuous as a
map from $\ovl{\A_0}$
to $\A$ (both equipped with their respective norm topologies). Hence it cannot
be extended to $\A_0$
in any natural way. The problem here is that we wish to work in a
Banach-algebraic framework; the
map $\Qh$ {\em is} continuous  as an operator from $L^2(T^*\Rn)$ to
$HS(L^2(\Rn))$ (if both are
regarded as Hilbert spaces, the latter being the space of Hilbert-Schmidt
operators on $L^2(\Rn)$,
with the inner product $(A,B)={\rm Tr} AB^*$), and also as a map from the
Schwartz space ${\cal
S}'(T^*\Rn)$ to the space of continuous linear maps from ${\cal S}(\Rn)$ to
${\cal S}'(\Rn)$, cf.\
\ci{Fol,GB}.

Both problems may be resolved simultaneously if we construct a positive
quantization, that is, find
a map $\Qh':\A_0\raw \A$ which is positive. For a positive map between two
$C^*$-algebras is
automatically continuous (see \ci{Tak}, p.\ 194). Let $\{\Oh\}_{\pl>0}$ be a
family of
normalized vectors in $L^2(\Rn)$, which satisfy the condition that in the limit
$\pl\raw 0$
 the Wigner function $W^{\pl}_{\Oh}$ is smooth in all variables (including
$\pl$), vanishes rapidly
at infinity, and converges   to $(2\pi)^n\dl(x,p)$ in the distributional
topology defined by the test
function space $\ovl{\A_0}$ (defined after (\ref{3.1})). An example is \be
\Oh(x)=(\pi\pl)^{-n/4}e^{-x^2/2\pl}, \ll{p3}
\ee
with Wigner function
\be
\wi_{\Oh}(x,p)=(2/\pl)^ne^{-(x^2+p^2)/\pl}. \ll{p4}
\ee
We then define a new quantization map $\Qh^{\Om}$ by
\be
\Qh^{\Om}(f)=\Qh(\til{\wi_{\Oh}}*f), \ll{p5}
\ee
 with $\Qh$ the Weyl quantization (\ref{3.3}), $\til{\wi_{\Oh}}$ defined by
$\til{\wi_{\Oh}}(x,p)=
\wi_{\Oh}(-x,-p)$, and $*$ being the convolution product in ${\Bbb R}^{2n}$. It
follows from Prop.\
1.99 in \ci{Fol} that $\Qh^{\Om}$ is a positive map. Since the uniform operator
norm is majorized by
the Hilbert-Schmidt norm, it follows from the triangle inequality and the first
continuity property
of $\Qh$ mentioned above that $\Qh^{\Om}$ defines a quantization if for all
$f,g\in\ovl{\A_0}$
\bea
L^2-\lpl \left(
\{\til{\wi_{\Oh}}*f,\til{\wi_{\Oh}}*g\}-\til{\wi_{\Oh}}*\{f,g\}\right)&=& 0;
\nn\\
L^2-\lpl \left( (\til{\wi_{\Oh}}*f)\cdot
(\til{\wi_{\Oh}}*g)-\til{\wi_{\Oh}}*(f\cdot g)\right)&=& 0,
\ll{p6} \eea
and if the function $\pl\raw \n \Qh^{\Om}(f)\n$ is continuous for all such
$f$. These conditions
are all satisfied if $\Oh$ is as specified prior to (\ref{p3}), and thus
$\Qh^{\Om}$ is indeed a
positive definite quantization (note that $ \Qh^{\Om}$ is automatically
self-adjoint, since
$\wi_{\Oh}$ is real-valued). It can be extended to $\A_0$ by continuity, but
the
extension obviously does not satisfy the quantization condition involving the
Poisson bracket (which
is not a continuous map on $\A_0$ in either variable).

This procedure may be extended to arbitrary manifolds $Q$; the smearing $f\raw
\til{\wi_{\Oh}}*f$
will in general be replaced by the use of Friedrichs mollifiers. It is clear
that this positive
definite quantization procedure is not intrinsic: it depends on the choice of
the $\Oh$. It may be
argued that the Weyl quantization procedure is not intrinsic either, because
from a geometric point
of view \ci{NPLstr} it relies on the choice of a diffeomorphism between a
tubular neighbourhood of
$Q$ in $TQ$, and one of $\Dl{Q}$ in $Q\times Q$. In any case, one may argue
that points in space
should be stochastic objects, with a probability distribution related to $\Oh$.
This point of view is
defended, in a quite different context, in \ci{Pru,Ali}.
\section{Lie groupoids, Lie algebroids, and their Jordan-Lie algebras}\eo
The (generalized \ci{NPLstr}) Weyl quantization of $C_0(T^*Q)$ by
$\K(L^2(Q))_{\rm sa}$ and
the quantization of $C_0(\g^*)$ by $JL(G)=C^*(G)_{\rm sa}$ are both special
cases of a rather
general construction involving Lie groupoids, which are a certain
generalization of Lie groups that
are of great physical and mathematical relevance (cf.\ \ci{Mac,CDW} for a
comprehensive discussion
of these structures, illustrated with many examples).
\subsection{Basic definitions}
We recall that a category $G$ is a class $B$ of objects together with a
collection of arrows. Each
arrow $x$ leads from object $s(x)$ (the source of the arrow) to the object
$t(x)$ (the target). If
$s(x)=t(y)$ then the composition $xy$ is defined as an arrow from $s(y)$ to
$t(x)$, and this partial
multiplication on $G$ is associative whenever it is defined. Also, each object
$b\in B$ comes with
an arrow $i(b)$, which serves as the identity map from $s(i(b))=b$ to
$t(i(b))=b$, so that $xi(b)=x$
(defined when $s(x)=b$) and $i(b)x=x$ (defined when $t(x)=b$). Hence we obtain
an inclusion $i$ of
$B$ into $G$. A category is called small if $B$ is a set.
\begin{defn}
A groupoid is a small category in which each arrow is invertible. \ll{groupoid}
\end{defn}
Hence for each $x\in G$ the arrow $x^{-1}$ is defined, with $s(x^{-1})=t(x)$
and $t(x^{-1})=s(x)$,
and one has $i\circ s(x)=x^{-1}x$ and $i\circ t(x)=xx^{-1}$. We may regard $G$
as a fibered space
over $B$, with two  projections $S:G\raw B$ and $t:G\raw B$. One may pass to
topological groupoids
by requiring continuity of the relevant structures, and to Lie groupoids by
demanding smoothness:
\begin{defn}
A Lie groupoid is a groupoid in which $G$ and $B$ are smooth manifolds (taken
to be Hausdorff,
paracompact and finite-dimensional), so that the inclusion $i$ is a smooth
embedding, the
projections $s$ and $t$ are smooth surjective submersions, and  the inverse
$x\raw x^{-1}$,  as
well as the partial multiplication $(x,y)\raw xy$ are smooth maps.
\ll{Lie}
\end{defn}
Variations on this definition are possible, cf.\ \ci{Mac,CDW}; for example, in
the former ref.\ the
assumption is added that $G$ is transitive, in the sense that the map $s\times
t:G\raw B\times B$ is
surjective (that is, any two points in $B$ can be connected by an arrow), but
since a corresponding
transitivity assumption is not part of the definition of a Lie algebroid (see
below) in \ci{Mac}, we
follow \ci{CDW} in dropping it.

We see that a Lie group is a special case of a Lie groupoid, namely a case in
which $B$ consists of
one point $b$ (and $i(b)=e$ is the identity of $G$), so that all arrows can be
composed. One may
generalize the passage from a Lie group to a Lie algebra in the present
context.
First note that each $x\in G$ defines not only an arrow from $s(x)$ to $t(x)$,
but in addition leads
to a map $L_x:t^{-1}(s(x))\raw t^{-1}(t(x))$, defined by $L_x(y)=xy$.
Similarly, one has a map
$R_x:s^{-1}(t(x))\raw s^{-1}(s(x))$ given by $R_x(y)=yx$. As in the Lie group
case, we would like to
define left- and right invariant vector fields on $G$. Hence we would obtain
(say) a left-invariant
flow $\phv_{\ta}$ on $G$, satisfying $\phv_{\ta}( L_x(y))=L_x \phv_{\ta}(y)$
for $y\in t^{-1}(s(x))$.
The problem is
that $L_x$ is only a partially defined multiplication, so that the right-hand
side is only defined
if $t(\phv_{\ta}(y))=s(x)$, that is, the target of $\phv_{\ta}(y)$ must not
depend on the time
$\ta$. Hence $(d/d\ta) t(\phv_{\ta})=0$, or $t_* X=0$ if $X=(d/d\ta)
(\phv_{\ta})_{|\tau=0}$.
In conclusion, we may define a left-invariant vector field $\xi_L$ by the
conditions
\be
t_*\xi_L=0;\:\:\: (L_x)_*\xi_L=\xi_L\:\: \forall x\in G, \ll{5.1}
\ee
and a right-invariant vector field $\xi_R$ by the conditions
\be
s_*\xi_R=0;\:\:\: (R_x)_*\xi_R=\xi_R \:\: \forall x\in G. \ll{5.2}
\ee
It is easily shown \ci{Mac,CDW} that the commutator (Lie bracket) of two left
(right) invariant
vector fields is left (right) invariant. Hence we may define
\begin{defn}
The Lie algebroid $\g$ of a Lie groupoid $G$ is the real vector space of all
vector fields on $G$
satisfying (\ref{5.1}), equipped with the following structures:\\
i) a projection $pr:\g\raw B$ (namely the obvious one, coming from the
projections $TG\raw
G\stackrel{s}{\raw} B$), which makes $\g$ a vector bundle over $B$;\\
ii) a projection $q:\g\raw TB$, given by $q=s_*$;\\
iii) a Lie bracket on $\Gm(\g)$ (the space of smooth sections of $\g$), which
is given by the
commutator on $\Gm(TG)$, and which satisfies
\be
q([\xi_{L}^{1},\xi_{L}^{2}])  =  [q(\xi_{L}^{1}),q(\xi_{L}^{2})];\ll{5.3a}\ee
\be
[\xi_{L}^{1},f\xi_{L}^{2}]  =  f[\xi_{L}^{1},\xi_{L}^{2}]+q(\xi_{L}^{1})f\cdot
\xi_{L}^{2}\:\:\:
 \forall f\in \cin(B). \ll{5.3}
\ee
\ll{algebroid}
\end{defn}
Of course, an equivalent definition is obtained by replacing (\ref{5.1}) by
(\ref{5.2}), and $s$
and $s_*$  by $t$ and $t_*$, respectively. One may define a Lie algebroid
without reference to Lie
groupoids as vector bundle $E$ over $B$, together with an additional projection
$q:E\raw TB$
satisfying (\ref{5.3a}) (the `anchor' of $E$ \ci{Mac}) and a Lie bracket on
$\Gm(E)$ satisfying the
analogue of (\ref{5.3}).  A Lie algebroid is called transitive if $q$ is a
surjective; if a Lie
groupoid $G$ is transitive then so is its algebroid $\g$.  A simple example is
$E=TQ$ as a vector
bundle over $Q$, with $q$ the identity map.

One may generalize the identification $\g\simeq T_eG$ for a Lie group $G$ as
follows. Since
$x=x(x^{-1}x)=L_x(i\circ s(x))$, every left-invariant vector field $\xi_L$ on
$G$ is determined by its
values at $i(B)\equiv G_0 \subset G$. We have the decomposition
$T_{G_0}G=T_{G_0}G_0\oplus {\rm
ker}(t_*)\lceil T_{G_0}G$ (where $\lceil$ means `restricted to'), so we see
that $\g\simeq
T_{G_0}G/T_{G_0}G_0$, which is just the normal bundle $N_i$ of the embedding
$i:B\raw G$.
Equivalently, if we define $T^tG$ as the vector bundle over $G$ consisting of
elements of $TG$
annihilated by $t_*$ (with the canonical projection $pr_t$
onto $G$  borrowed from $TG$), then $\g=i^*(T^tG)$, the pull-back bundle over
$B$ given by the map
$i:B\raw G$. Conversely, $T^tG=s^*(\g)$ as a pull-back bundle \ci{Mac}. Note
that $T^tG$ is itself a
Lie algebroid over $G$, with the anchor $q:T^tG\raw TG$ just given by
inclusion.

An interesting property of a Lie algebroid $E$  is that a connection on $E$
allows one to define
generalized geodesics on $E$ (hence on the base space $B$). Namely, one obtains
a vector field $\xi$
on $E$, whose value at $Y\in E$  is given by the horizontal lift of $q(Y)\in
TB$ at $Y$. The flows
of this field are the desired generalized geodesics (for $E=TB$ equipped with
the Levi-Civita
connection these are the usual geodesics). This leads to the construction of a
map $\exp:\g\raw G$
which generalizes the one for Lie groups \ci{Pra}. Namely, by the preceding
paragraph $T^tG$
regarded as a vector bundle over $G$ inherits  the chosen connection $A$ on
$\g$ (considered a vector
bundle over $B$) as the pull-back $s^*A$, and this implies that one has a
generalized geodesic flow
$\gm_{\ta} $ on $T^tG$. Now for $X\in\g$,
\be
e^X=pr_t(\gm_1(X)), \ll{5.4}
\ee
where on the right-hand side we regard $X\in\g\subset T^tG$ via the natural
embedding of $\g
 \equiv {\rm
ker}(t_*)\lceil T_{G_0}G$ in $T^tG$. If $G$ is a Lie group then obviously no
connection needs
to be chosen (all vectors on $\g$ are vertical, so the geodesic flow on $\g$ is
the identity map),
and the map $\exp$ reduces to the usual one.
\subsection{Algebras of observables from Lie algebroids and groupoids}
Generalizing the Poisson algebra $\cin(\g^*)$ of a Lie algebra $\g$ (which is a
vector bundle over a
single point), one may associate a Poisson algebra $\cin(E^*)$ to any Lie
algebroid $E$ \ci{CDW}; here
$E^*$ is the dual of $E$ as a vector bundle.    The Poisson structure is
completely
determined by specifying the Poisson bracket between   arbitrary sections
$\xi_1,\xi_2$ of $E$ and functions $f_1,f_2$ on $B$. Here any $\xi\in\Gm(E)$
defines an element
$\til{\xi}\in\cin(E^*)$ as follows: if $pr$ is the projection in $E^*$ then
$\til{\xi}(\th)=\langle
\th,\xi(pr(\th))\rangle$. These functions $\til{\xi}$ are obviously linear on
the fibers of $E^*$.
Furthermore, $f\in \cin(B)$ defined $\til{f}\in \cin(E^*)$ by pull-back.
The  Poisson brackets are
 \be
\{\til{\xi}_1,\til{\xi}_2\}=\widetilde{[\xi_1,\xi_2]}; \:\:\:
\{\til{f}_1,\til{f}_2\}=0;\:\:\:
\{\til{\xi},\til{f}\}=\widetilde{q(\xi)f}. \ll{5.5} \ee
This bracket may
subsequently be extended to a dense subset of $\cin(E^*)$ (in a suitable
topology) by imposing the
Leibniz rule on products of linear functions. On $E=TQ$ this procedure is
equivalent to imposing the identities
$\{\sg(\xi_1),\sg(\xi_2)\}=\sg([\xi_1,\xi_2])$,
$\{\til{f}_1,\til{f}_2\}=0$, and $\{\sg{\xi},\til{f}\}=\widetilde{\xi(f)}$,
where
$\sg(\xi)\in \cin(T^*Q)$ is the symbol of the vector field $\xi$ on $Q$. This
leads to the
canonical Poisson structure on $\cin(T^*Q)$.  In case that $E=\g$ is the Lie
algebroid of a Lie
groupoid, a more intrinsic construction of this Poisson structure is given in
\ci[II.4.2]{CDW}.

In similar spirit, we can construct a non-commutative $C^*$-algebra (hence a
non-associative
Jordan-Lie algebra) from a Lie groupoid $G$ (indeed, from almost any
topological groupoid \ci{Ren},
but the construction is more canonical in the Lie case, where a natural measure
class is singled out,
see below). To do so, we need to chose a measure $\mu_b$ on each fiber
$t^{-1}(b)$ of $G$, in such a
way that the family of measures thus obtained is left-invariant (that is, the
map
 $L_x:t^{-1}(s(x))\raw t^{-1}(t(x))$ should be measure-preserving for all $x$).
Since the fibers are
manifolds, we naturally require that each measure $\mu_b$ is equivalent to the
Lebesgue measure (on
a local chart). The precise choice of the $\mu_b$ does not matter very much in
that case, as the
$C^*$-algebras corresponding to different such choices will be isomorphic. In
both the Lie and the
general case, groupoid $C^*$-algebras are of major mathematical interest, as
they provdide
fascinating examples of non-commutative geometry (cyclic cohomology) and
topology (K-homology), cf.\
\ci{Con82,MS}. The algebra is constructed starting from $\cci(G)$, which is
equipped with a product
\be
f*g(x)=\int_{t^{-1}(s(x))}d\mu_{s(x)}(y)\, f(xy)g(y^{-1}), \ll{5.6}
 \ee
and an involution
\be
f^*(x)=\ovl{f(x^{-1})}, \ll{5.7}
\ee
  which are clearly generalizations of (\ref{3G1}). The construction of the
norm is described in
\ci{Ren} (for general  groupoids), and the closure of $\cci(G)$ in this norm is
the groupoid algebra
$C^*(G)$. Its self-adjoint part, with the multiplications $\sh$ and $\ah$ (cf.\
(\ref{3.5})), is the
Jordan-Lie algebra $JL(G)$.

For $G$ a Lie group we thus recover the group algebra, whose representation
theory is discussed in
subsect.\ 3.4;  the opposite case is the so-called coarse groupoid $G=Q\times
Q$, where $Q$ is a
manifold.  This has base space $B=Q$, and source and target projections
$s((x,y))=y$, $t((x,y))=x$.
The inclusion is $i(x)=(x,x)$, the inverse is $(x,y)^{-1}=(y,x)$, and the
composition rule is
$(x_1,y)(y,x_2)=(x_1,x_2)$. The measures $\mu_b$ may all be taken to be
identical to a single measure
$\mu$ on $Q$, and one easily finds that $C^*(Q\times Q)=\K(L^2(Q;\mu))$, cf.\
\ci{Con82}. Its
self-adjoint subspace $JL(Q\times Q)$ is the quantum algebra of observables of
a particle moving on
$Q$ \ci{NPLstr}, and it will not come as a surprise that the Poisson algebra of
the Lie algebroid
$TQ$ of $Q\times Q$ is just $\cin(T^*Q)$, the classical algebra of observables
of the particle. The
quantum algebra  $JL(Q\times Q)$ has only one irreducible representation,
namely the defining one on
$L^2(Q;\mu)$  (up to unitary equivalence). Similarly,  the classical algebra
$\cin(T^*Q)$ has only
one classical irreducible representation (up to symplectomorphisms), given by
$S=T^*Q$. These are
the quantum as well as classical Jordan-Lie analogues of the well-known Stone-
von Neumann
uniqueness theorem on regular representations of the canonical commutation
relations (see e.g.\
\ci{BR2,Fol,CG} for this theorem in its various settings).

The situation where $G$ is either a Lie group, or the coarse groupoid of some
manifold, are both
special cases of so-called gauge groupoids \ci{Mac,CDW}. A gauge groupoid is
equivalent to a
principal fibre bundle $(P,Q,H)$, where $P$ is the total space, $Q$ is the base
space, and $H$ is a
Lie group acting on $P$ from the right. The corresponding groupoid is denoted
by $P\times_H P$. It
is a quotient of the coarse groupoid $P\times P$, obtained   by imposing the
equivalence relation
$(x_1,x_2)\sim (y_1,y_2)$ iff
$(x_1,x_2)= (y_1h,y_2h)$ for some $h\in H$; we denote the equivalence class of
$(x,y)$ by $[x,y]$.
Accordingly, $B=Q=P/H$, the inverse is $[x,y]^{-1}=[y,x]$, the projections are
$s([x,y])=pr_{P\raw
Q}(y)$,  $t([x,y])=pr_{P\raw Q}(x)$, the inclusion is $i(q)=[s(q),s(q)]$ (for
an arbitrary section
$s$ of $P$), and  multiplication $[x_1,y_1]\cdot [y_2,x_2]$ is defined iff
$y_2=y_1h$ for some $h\in
H$, and the composition equals $[x_1h,x_2]$ in that case. For $H=\{e\}$ we get
the coarse groupoid,
and for $P=H=G$ we get a Lie group $G$. It can be shown that any transitive
groupoid is of the form
 $P\times_H P$ \ci{Mac}.

 If $H$ is compact the groupoid $C^*$-algebra is $C^*( P\times_H P)\simeq
C^*(Q\times
Q)\ot C^*(H)$ \ci{NPLstr}, which is the quantum algebra of observables of a
particle moving on $Q$
with an internal degeree of freedom, namely  a charge coupling to a gauge field
defined on the
bundle $(P,Q,H)$. The Lie algebroid of  $P\times_H P$ is $(TP)/H$ (where the
action of $H$ on $TP$
is the push-forward of its action on $P$). The corresponding Poisson algebra
$\cin((T^*P)/H)$ was
already known to be the classical algebra of observables of a particle coupling
to a Yang-Mills
field \ci{GS}, and it is satisfying that the quantum algebra $C^*( P\times_H
P)_{\rm sa}$ can be
obtained as a deformation of it; the quantization maps $\Qh$ are given in
\ci{NPLstr}.

The irreducible representations of the classical algebra of observables
$A_0=\cin((T^*P)/H)$
correspond to the  symplectic leaves of $(T^*P)/H$ (and their covering spaces),
which are discussed
in \ci{GS}. There is a one-to-one correspondence between the set of these
leaves, and the set of
co-adjoint orbits in $\h^*$ (the dual of the Lie algebra of $H$): each leaf
$P_{\O}$ is a fiber
bundle over $T^*Q$ whose characteristic fiber is the co-adjoint orbit $\O$.
Hence locally
$P_{\O}\simeq T^*Q\times \O$, and the orbit $\O\subset \h^*$ clearly serves as
a classical internal
degree of freedom (`charge') of the particle. Hence the representation theory
of  $\cin((T^*P)/H)$ is
isomorphic to that of $\cin(\h^*)$ with the Lie etc. Poisson structure
discussed in subsect.\ 2.3.

An analogous situation prevails in the quantum case $\A=JL( P\times_H P)$
\ci{NPLstr}. The
representation theory of this algebra is isomorphic to that of $JL(H)$ (see
subsect.\ 3.4), hence
each irreducible unitary representation $\pi_{\ch}$  of $H$ on a Hilbert space
$\Hs$ induces an
irreducible representation $\pi^{\ch}$ of $\A$, and {\em vice versa}. The
Hilbert space $\H^{\ch}$
carrying the representation $\pi^{\ch}(\A)$ is naturally realized as
$\H^{\ch}\simeq L^2(Q)\ot
\H_{\ch}$, so that we see that $\Hs$ acts as an internal degree of freedom of
the particle (a
`quantum charge').

To sum up, we see that classical internal degrees of freedom are co-adjoint
orbits of a Lie group,
whereas the quantum analogue of this is an irreducible unitary representation
of the same group,
compare with the discussion in subsect.\ 3.4.

 We end in a speculative manner. In \ci{NPLstr}
 one finds a proof of the transitive case of the following
\begin{con}
 Let $G$ be a Lie groupoid, and $\g$ its Lie algebroid. Then there exists a
quantization relating
the Poisson algebra $\cin(\g^*)$ canonically associated to $\g$  to the
Jordan-Lie algebra
$JL(G)=C^*(G)_{\rm sa}$, in the sense of Definition \ref{qua}.
\end{con}
 
\end{document}